\documentclass[fleqn,twoside]{article}
\usepackage{espcrc2}

\usepackage{graphicx}
\usepackage{amsfonts, color, float, bbm, amsmath}
\usepackage{amssymb,amsthm}

\def\beq {\begin{equation}}
\def\eeq {\end{equation}}
\def\bea {\begin{eqnarray}}
\def\eea {\end{eqnarray}}

\def\ni {\noindent}
\def\nn {\nonumber}

\def\ie{{\it i.e.}}
\def\eg{{\it e.g.}}

\title{Duality violations in $\tau$ hadronic spectral moments}

\author{D. R. Boito,\thanks{Speaker}\address[UAB]{Grup de F\'{\i}sica Te\`orica, Universitat Aut\`onoma de Barcelona,
E-08193 Bellaterra (Barcelona), Spain}$^,$\address[IFAE]{Institut de F\'isica d'Altes Energies (IFAE), Campus UAB, E-08193 Bellaterra (Barcelona), Spain}
        O. Cat\`a,\address{Departament de F\'isica Te\`orica and IFIC, Universitat de Val\`encia-CSIC,
\\Apt. Correus 22085, E-46071 Val\`encia, Spain}
        M. Golterman,\address[Maartens]{Department of Physics and Astronomy, San Francisco State University, San Francisco, CA 94132, USA}
        M. Jamin,\addressmark[IFAE]$^,$\address{Instituci\'o Catalana de Recerca i Estudis Avan\c cats (ICREA)}
        K. Maltman,\address{Math and Statistics, York University, 4700 Keele ST.
Toronto, ON Canada}$^,$\address{CSSM, University of Adelaide, Adelaide, SA
Australia}
        J. Osborne,\addressmark[Maartens] and
        S. Peris\addressmark[UAB]
}
       
\begin{document}

\begin{abstract}

Evidence is presented for the necessity of including duality violations in a
consistent description of spectral function moments employed in the precision
determination of $\alpha_s$ from $\tau$ decay. A physically motivated
{\em ansatz} for duality violations in the spectral functions enables us to
perform fits to spectral moments employing both pinched and unpinched weights.
We describe our analysis strategy and provide some preliminary findings. Final
numerical results await completion of an ongoing re-determination of the ALEPH
covariance matrices incorporating correlations due to the unfolding procedure
which are absent from the currently posted versions. To what extent this issue
affects existing analyses and our own work will require further study.
\vspace{1pc}
\end{abstract}

\maketitle

\section{Introduction}

Hadronic decays of the $\tau$ lepton provide a particularly clean environment
for the study of low-energy QCD. The mass of the $\tau$ (which is the only
lepton heavy enough to decay into hadrons) is large enough that perturbative
QCD should provide a good description of the inclusive decay process, but not
so large that nonperturbative effects can be completely neglected.
  
Since the seminal work of Ref.~\cite{BNP1992}, it has been recognised that
the perturbative series supplemented with the Operator Product Expansion (OPE)
provides a solid framework for describing these decays. The main achievement
of this program is a precision determination of the QCD coupling $\alpha_s$.
The determination of $\alpha_s$ at different scales provides a highly nontrivial
test of the theory. Hadronic $\tau$ decays also yield values for vacuum
condensates~\cite{ALEPH}, low-energy constants~\cite{PichetalDVs},
and the CKM matrix element $|V_{us}|$~\cite{Vus1}.

Unfolded high-statistics spectral functions extracted from the final LEP data
are available from ALEPH~\cite{ALEPH} and OPAL~\cite{OPAL}. On the theory side,
the $\mathcal{O}(\alpha_s^4)$ term of the perturbative contribution has been
calculated recently~\cite{Baikovetal2008}. This has triggered several reanalyses
of hadronic $\tau$ decays from the ALEPH and OPAL spectral functions. While the
average value of $\alpha_s$ from $\tau$ data is competitive with the current
world average, the values obtained by different groups are barely compatible.
After the evolution to the $Z$ mass, a sample of results obtained by various
groups is
\begin{equation}
\alpha_s(m_Z^2) =\begin{cases}
\;0.1202\, (6)_{\rm exp} (18)_{\rm 
th}\,\,\,\,\,\mbox{\cite{Baikovetal2008}}\,\ ,\cr
\;0.1212\, (5)_{\rm exp} (9)_{\rm th}\,\,\,\,\,\ 
\,\mbox{\cite{Davieretal2008}}\,\
,\cr
\;0.1180\, (4)_{\rm exp} (7)_{\rm th}\,\,\,\,\, \ 
\,\mbox{\cite{BenekeJamin2008}}\
\,, \cr
\;0.1187\, (6)_{\rm exp} (15)_{\rm 
th}\,\,\,\,\,\mbox{\cite{MaltmanYavin2008}}\
\,.
\end{cases}
\label{alphas}
\end{equation}
Furthermore, there is some tension between the result of
Ref.~\cite{Davieretal2008} and recent lattice determinations; \eg, the
analysis of Ref.~\cite{LatticeAlphaS} gives
\begin{equation}
\alpha_s(m_Z^2) = 0.1184 \pm 0.0006.
\end{equation}
Even though much of the variation in Eq.~(\ref{alphas}) results from
differences in the prescriptions used for resumming the perturbative series
(CIPT versus FOPT~\cite{Pivovarov,BenekeJamin2008,CapriniFischer}), at the
current level of precision nonperturbative effects thus far neglected may
become significant.

One can think of several different, but not completely independent, sources of
systematic uncertainties in the theoretical description of hadronic $\tau$
decays.  Examples are the truncation in powers of $\alpha_s$ and/or variation
in the choice of resummation prescription for the perturbative series
\cite{BenekeJamin2008,CapriniFischer}, unchecked assumptions regarding the
truncation in dimension of the OPE \cite{MaltmanYavin2008}, and contributions
from duality violations (DVs)~\cite{Shifman,CGP2005}. In this work we address
the latter. The OPE is believed to be an asymptotic expansion at best, and it
breaks down on the Minkowski axis. This lack of convergence is related to the
presence of DVs, and indeed, our preliminary results confirm previous
observations on the presence of nonnegligible DVs for certain weighted spectral
integrals~\cite{km98}.  A full reanalysis including all systematic effects
noted above will be presented in a forthcoming article.

%%%%%%%%%%%%%%
\section{Theoretical framework }

We start from the total nonstrange branching ratio $R_\tau^{S=0}$ defined as
\begin{equation}
R_\tau^{S=0} = \frac{\Gamma[\tau \to \mbox{hadrons}^{S=0}\,
\nu_\tau]}{\Gamma[\tau \to e^- \bar \nu_e\,
\nu_\tau]}=R_\tau^V+R_\tau^A\,.
\end{equation}
With vector ($V$) and axial-vector ($A$) currents $J^V_\mu(x)= \bar u
\gamma_\mu d(x)$ and $J^A_\mu(x)= \bar u \gamma_\mu\gamma_5 d(x)$, the
$J=0,\,1$ parts of the $V$ and $A$ current two-point functions,
$\Pi^{(J)}_{V,A}$, are defined by
\begin{eqnarray}
\Pi^{V,A}_{\mu\nu}(q) \!\!\!\!&=&\!\!\!\! i \!\int d^4 x \,e^{i qx}\langle
0|T\{J^{V,A}_\mu(x) J^{V,A}_\nu(0)^\dagger\}|0\rangle \nn \\
&&\hspace{-1.8cm}=(q^{\mu}q^{\nu}- q^2 g^{\mu\nu})\Pi_{V,A}^{(1)}(q^2)
+\,q^{\mu}q^{\nu}\Pi_{V,A}^{(0)}(q^2)\ .
\end{eqnarray}
The corresponding spectral functions
$\rho^{(J)}_{V,A}={\frac{1}{\pi}}\, {\rm Im}\, \Pi^{(J)}_{V,A}$ can be
extracted from the differential distribution $dR_\tau^{V,A}/dq^2$, which is
experimentally available~\cite{ALEPH,OPAL}. Explicitly, with $s=q^2$,
\begin{eqnarray}
{\frac{dR_\tau^{V,A}}{ds}} \!\!\!\!&=&\!\!\!\! 12\pi^2 S_{\rm{EW}} |V_{ud}|^2
{\frac{1}{m_\tau^2}} \left(1-\frac{s}{m_\tau^2}\right)^2 \times \nn\\ 
&&\hspace{-1.1cm}\left[\left(1+2\frac{s}{m_\tau^2}\right)\rho_{V,A}^{(1+0)}(s)-
2\frac{s}{m_\tau^2}\rho_{V,A}^{(0)}(s) \right]\,.
\label{Rtau}
\end{eqnarray}
Except for the pion-pole contribution to $\rho_A^{(0)}(s)$, the $J=0$ part
of the spectral function is numerically negligible. The combination
$\Pi_{V,A}(s)\equiv\Pi_{V,A}^{(1+0)}(s)$ is free of kinematic singularities
and analytic in the complex $s$-plane cut along the positive real axis.
Cauchy's theorem thus yields the finite-energy sum rule (FESR)\footnote{In
the remainder the factor of $ S_{\rm{EW}} |V_{ud}|^2$ is absorbed into
$\Pi_{V,A}(s)$.}
\begin{eqnarray}
R^{[w]}_{V,A}(s_0) \!\!\!\!&=&\!\!\!\! 12\pi^2 \int\limits_{0}^{s_0}
\frac{ds}{s_0}\, w(s/s_0)\,\rho_{V,A}(s) \nn\\
\!\!\!\!&=&\!\!\!\! 6\pi i \!\!\oint\limits_{|s|=s_0}\frac{ds}{s_0}\,
w(s/s_0)\, \Pi_{V,A}(s)\,, 
\label{sumrule1}
\end{eqnarray}
valid for an arbitrary analytic weight $w(z)$. 

For $s_0\gg\Lambda^2_{\rm QCD}$, one can expect $\Pi_{V,A}$ to be well
approximated by the OPE, \ie, 
\begin{equation}
\label{DeltaDVs}
\Pi_{V,A}(s) = \Pi_{V,A}^{\rm OPE}(s) + \Delta_{V,A}(s)\,,
\end{equation}
with $\Delta_{V,A}(s)$ a small (but in general nonzero) correction accounting
for the presence of DVs. 

Assuming DVs to vanish sufficiently fast for $|s|\to \infty$ in the whole
complex plane, one can show that the correction from $\Delta_{V,A}(s)$ to the
contour integral in Eq.~(\ref{sumrule1}) can also be written as~\cite{CGP2008}
\begin{equation}
\mathcal{D}^{[w]}_{V,A}(s_0) = -\, 12\pi \int\limits_{s_0}^{\infty}
\frac{ds}{s_0}\, w(s/s_0)  \,{\rm Im}\,\Delta_{V,A}(s)\,.
\label{DVsCorr}
\end{equation}
FESRs thus provide constraints on ${\rm Im}\,\Delta_{V,A}(s)$ beyond those
from the region $s_0<s<m_\tau^2$ obtained by fitting to the experimental
spectral function. In standard $\tau$-decay-based determinations of $\alpha_s$,
$\Delta_{V,A}(s)$ is typically neglected, with at most a check on the
self-consistency of this assumption~\cite{MaltmanYavin2008}.\footnote{An
exception is Ref.~\cite{Davieretal2008}, which discusses the relevance of this
term, though only for the combined $V+A$ correlator, and concludes that its
contribution lies within the error bars. On the basis of the analysis of
Ref.~\cite{CGP2008} and the present study we believe this conclusion needs to
be reconsidered.} It is our aim to quantitatively determine the impact of DVs
on the values for the QCD parameters accessible through such fits.

Since at present no theory for DVs exists, we adopt the following {\em ansatz}
for the asymptotic form of the DVs as a working hypothesis~\cite{CGP2008}
\begin{equation}
\frac{1}{\pi}{\rm Im}\,\Delta_{V,A}(s)\underset{\rm{large}\, s}\longrightarrow \kappa_{V,A} e^{-\gamma_{V,A} s}\sin(\alpha_{V,A}+\beta_{V,A}s)\,,
\label{DVsAnsatz}
\end{equation}
reflecting the presumed asymptotic character of the OPE expansion. Model
studies suggest the exponential decay, originating from the finite width of
the resonances. The oscillatory behaviour arises naturally in a spectral
function with resonances distributed with some periodicity as, for instance,
on the daughter trajectories in Regge theory~\cite{Shifman,CGP2005,CGP2008}.
With the (eight) parameters $\kappa_{V,A}$, $\gamma_{V,A}$, $\alpha_{V,A}$,
and $\beta_{V,A}$  we expect the {\em ansatz} of Eq.~(\ref{DVsAnsatz}) to
capture the generic features of the DVs, thus avoiding any more specific
models. This {\em ansatz} is certainly more general, and hence more likely
to be realistic, than that used in analyses which entirely neglect DV
contributions (which correspond to the special case $\kappa_V=\kappa_A=0$).

With our description of ${\rm Im}\, \Delta_{V,A}(s)$, the experimental $V$ and
$A$ spectral functions constrain the DV parameters, provided the asymptotic
behaviour of Eq.~(\ref{DVsAnsatz}) has set in for $s>s_{min}$ with
$s_{\rm min}<m_\tau^2$. Above $s_{min}$ one then has 
\begin{eqnarray}
\rho_{V,A}(s) \!\!\!\!&=&\!\!\!\! \theta\left(s-s_{\rm min}\right)
\bigg[\, \frac{Nc}{12\pi^2} \left[1+\hat \rho(s) \right]  \nn\\
&&\hspace{-4mm} + \, \kappa_{V,A} e^{-\gamma_{V,A} s}\sin(\alpha_{V,A}+
\beta_{V,A}s) \,\bigg]\,.
\label{SpecFunc}
\end{eqnarray}
The function $\hat\rho(s)$ contains the perturbative corrections up to and
including $\mathcal{O}(\alpha_s^4)$ as well as condensate contributions. The
latter were  shown to be numerically irrelevant for
$s_{\rm min}\gtrsim 1.1$~GeV$^2$~\cite{CGP2008}. In Ref.~\cite{CGP2008}, fits
to ALEPH $V$ and $A$ spectral functions employing Eq.~(\ref{SpecFunc}) have
been carried out, testing the onset of the asymptotic behaviour of
Eq.~(\ref{DVsAnsatz}) in real data. The results show the model description of
${\rm Im}\,\Delta_{V,A}(s)$ to be reasonable, and certainly compatible with
ALEPH data. Similar tests confirm its compatibility with the OPAL data.

%%%%%%%%%%
\section{Strategy of our analysis}

With the OPE plus DV representation for the relevant correlators,
Eq.~(\ref{DVsCorr}) takes the form
\begin{eqnarray}
R^{[w]}_{V,A}(s_0) \!\!\!\!&=&\!\!\!\! 6\pi i \!\!\oint\limits_{|s|=s_0}
\frac{ds}{s_0}\, w(s/s_0)\, \Pi^{\rm OPE}_{V,A}(s) \nn\\ 
\!\!\!\!&\phantom{=}&\!\!\!\! +\, \mathcal{D}^{[w]}_{V,A}(s_0) \,.
\end{eqnarray}
The LH side is to be evaluated using $\rho_{V,A}^{(J)}$ extracted from the LEP
experimental data, while the RH side contains the various OPE parameters as
well as the parameters of the DV {\em ansatz}. Because of the high quality of
the spectral data, use of a range of $s_0$ with $s_0>s_{min}$ and a set of
weight functions $w_i(x)$ allow fits to obtain both OPE and DV parameters.

Different weight functions $w(x)$ emphasise different terms of the OPE of
$\Pi_{V,A}(s)$. For want of a good description of DVs, existing analyses have
been restricted to the case of so-called pinched weights (polynomials in
$x=s/s_0$ having zeros at $s=s_0$ which suppress contributions to the RHS of
Eq.~(\ref{sumrule1}) from the vicinity of the timelike point on the contour
$\vert s\vert =s_0$, where DVs are expected to be largest). Polynomials of
higher degree, which provide more ``pinching,'' are sensitive to higher order
terms in the OPE.  In some analyses, for practical reasons, the OPE has been
truncated at dimensions below that for which non-$\alpha_s$-suppressed
contributions are in principle present. This can introduce an uncontrolled
systematic uncertainty \cite{MaltmanYavin2008,CGP2008}.

The three main features of our strategy are as follows. First, because we
include DV contributions, our sum rules need not be restricted to pinched
weights only. Low-degree unpinched weights, sensitive to only a small number
of condensates (modulo $\alpha_s$-suppressed logarithmic corrections) are
employed together with pinched ones. Second, Eqs.~(\ref{DVsCorr}) and
(\ref{SpecFunc}) allow simultaneous fits to both the spectral function and the
moments $R^{[w]}_{V,A}(s_0)$. Third, following Ref.~\cite{MaltmanYavin2008},
we do not restrict ourselves only to $s_0=m_\tau^2$. The use of a range of
$s_0$ values allows for a better use of the available data, facilitating the
separation of condensate contributions of different dimension, and providing
a consistency check on the fits from which $\alpha_s(m_\tau^2)$ and the
condensates are extracted. Values for all fitted parameters should be
independent, within errors, of the precise set of $s_0$ and weight functions
used in their determination. We emphasise that this check has never been
performed in full before.\footnote{In Ref.~\cite{MaltmanYavin2008}, a window
of $s_0$ values {\it was} used. The restriction to $s_0>2\ {\rm GeV}^2$,
however, imposed by requiring the neglect of DVs to be self-consistent, made
it impossible to fit the gluon condensate $\langle \alpha_s G^2\rangle$, which
was therefore taken as external input. The strong anticorrelation found between
the fitted $\alpha_s$ and input $\langle \alpha_s G^2\rangle$ creates a
systematic uncertainty which can only be avoided in a framework like ours,
which includes DVs, and allows also $\langle \alpha_s G^2\rangle$ to be
obtained through a fit to data.}

In our framework, in order to determine the various DV parameters, one must
perform a separate analysis of $V$ and $A$ channels. With the eight parameters
of Eq.~(\ref{DVsAnsatz}) under control, one can then also analyse $V+A$ or
$V-A$. This allows us to consider the first Weinberg sum rule. The fact that
our fits satisfy this constraint provides a consistency check on the $V$ and
$A$ fits.

A comment is in order regarding an important technical issue. In real life,
data for the spectral functions are available in the form of binned histograms
and the spectral integrals in Eq.~(\ref{sumrule1}) are approximated by a finite
sum over bins. Even in the absence of correlations in the data, two different
moments $R^{[w]}_{V,A}(s_{1})$ and $R^{[w]}_{V,A}(s_{2})$ with $s_2>s_1$ will
be correlated since they share contributions from the bins $k$ with $s_k<s_1$.
This  correlation is stronger if $s_2$ and $s_1$ are closer to each other. In
practice, if one chooses adjacent bins, the correlation can be well over $90\%$.
Such strong correlations may generate very small eigenvalues in the correlation
matrix for the moments $R^{[w]}_{V,A}(s_i)$. This renders the task of
constructing reliable fits and a trustworthy treatment of uncertainties rather
involved. Nevertheless, strategies to deal with such difficulties
exist~\cite{MILC,Bohmetal}, and can be used to study the quality of the fits.

%%%%%%%%
\section{Evidence for duality violations}

To illustrate our argument, let us concentrate on fits using the weights
$w_1(x)=1$, $w_2(x)=1-x$, and $w_3(x)=(1-x)^2$. Since $w_1$ is unpinched, its
moments $R^{[w_1]}_{V,A}$ are expected to be more sensitive to DVs. The weights
$w_2$ and $w_3$ are singly and doubly pinched, respectively. The results
discussed in this section come from fits using 52 values of $s_0$ ranging from
$1.5125$~GeV$^2$ up to $2.7875$~GeV$^2$. In the perturbative contribution we
employ the Fixed Order prescription for the RG resummation.

\begin{figure}[!ht]
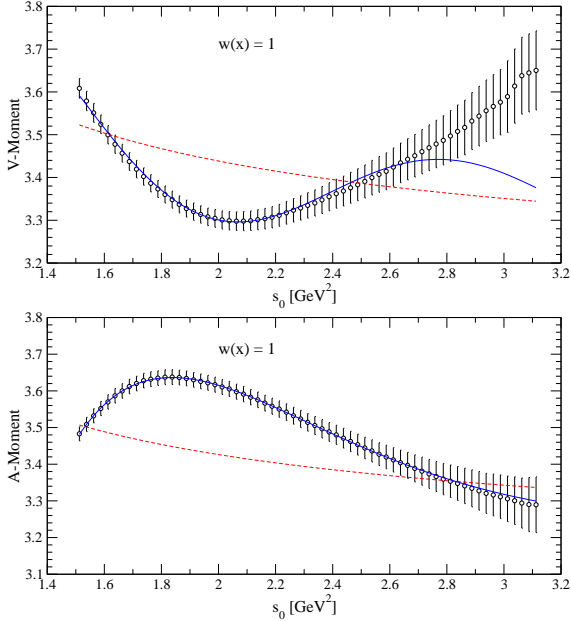

%\vspace{9pt}
%\framebox[55mm]{\rule[-21mm]{0mm}{43mm}}
\includegraphics[width=1\columnwidth,angle=0]{Vmom_w=1}
\includegraphics[width=1\columnwidth,angle=0]{Amom_w=1}
\vspace{-1.2cm}
\caption{Fits to the $w_1(x)=1$ spectral integrals for the $V$ channel (top)
and $A$ channel (bottom). The blue (solid) lines show fits including DVs
whereas the red (dashed) lines represent the model without DVs
($\kappa_{V,A}=0$).\label{w1}}
\end{figure}

In the fits using $w_1(x)$  the parameters are $\alpha_s(m_\tau^2)$ and the
eight DV parameters $\kappa_{V,A}$, $\gamma_{V,A}$, $\alpha_{V,A}$ and
$\beta_{V,A}$. (This moment is sensitive to the gluon condensate only through
tiny logarithmic corrections.) A comparison of the fit results for $w_1(x)$
with the experimental moments from ALEPH data is shown in Fig.~\ref{w1}. The
main feature of this figure is that a model in which DVs are neglected
($\kappa_{V,A}=0$, red line in Fig.~\ref{w1}) cannot account for the data. The
results obtained with DVs (blue line), on the other hand, compare very well
with the data. For the $A$ channel this is true right up to the kinematic
endpoint. For the $V$ channel, the agreement is less good above $~2.8$~GeV$^2$.
Here one should bear in mind that (1) high-statistics measurements of $4\pi$
modes from BaBar suggest the ALEPH $V$ spectral function may be overestimated
towards the end of the spectrum~\cite{4pi}, and (2) the OPE plus DV fit to the
OPAL $V$ channel data displays no such problem. This point will have to be
clarified in the future.

\begin{figure}[!ht]
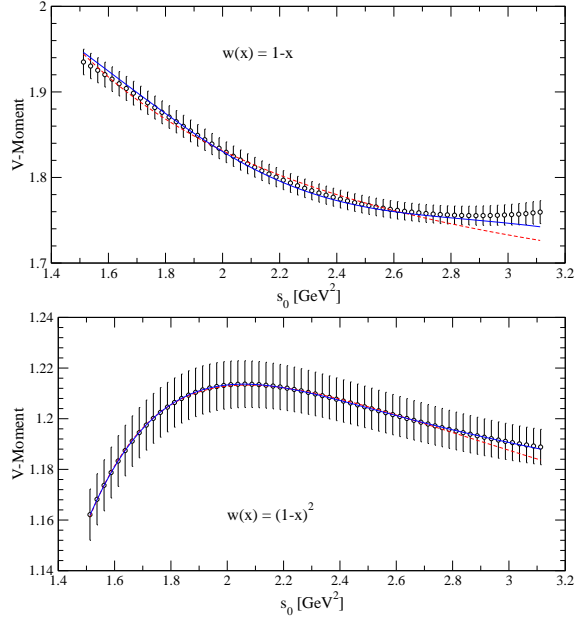

%\vspace{9pt}
%\framebox[55mm]{\rule[-21mm]{0mm}{43mm}}
\includegraphics[width=1\columnwidth,angle=0]{Vmom_w=1-x}
\includegraphics[width=1\columnwidth,angle=0]{Vmom_w=p1-xp2}
\vspace{-1.2cm}
\caption{Fits to the $V$ channel $w_2$ (top) and $w_3$ (bottom) spectral
moments. The blue (solid) lines show fits including DVs whereas the red
(dashed) lines represent the model without DVs ($\kappa_V=0$).\label{w1andw2}}
\end{figure}

Results for $V$ channel fits using the $w_2$ and $w_3$ moments are shown in
Fig.~\ref{w1andw2}. In these fits, the value of the gluon condensate is fitted
as well. Additionally, $w_3$ allows for a fit of the  $D=6$ condensate. The
fits with DVs (blue lines) are in all cases excellent. For $w_2$, the fit
including DVs is superior to that without. While for $w_3$ both fits are of
excellent quality, we recall that the fit without DVs requires
$\langle \alpha_s G^2\rangle$, which can only be determined in a fit which
includes DVs as input.

From such fits, values for  $\alpha_s$, the condensates and the DV parameters
can in principle be extracted. For the time being, we have not carried through the final version
of this analysis because of an inconsistency we uncovered in the correlation
matrix publicly available from ALEPH~\cite{ALEPHTables}. This problem is
explained in the next section.

%%%%%%%%%%%%%%%
\section{Correlations in the ALEPH spectral function data}

\begin{figure}[!ht]
%\vspace{9pt}
%\framebox[55mm]{\rule[-21mm]{0mm}{43mm}}
\vspace{-3mm}
\includegraphics[width=1\columnwidth,angle=0]{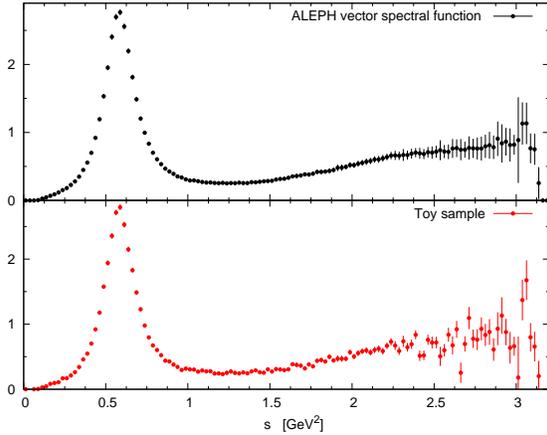}
\vspace{-1.2cm}
\caption{ALEPH data~\cite{ALEPHTables} for the vector spectral function (top)
and a MC toy data sample based on the ALEPH correlation matrix (bottom).
\label{ALEPH}}
\vspace{-0.5cm}
\end{figure}

In order to study the uncertainties associated with our fits, a Monte
Carlo generator of toy data sets was built based on ALEPH's spectra
and correlations~\cite{ALEPHTables}.\footnote{We thank A.~H\"ocker for
  this suggestion.} The unfolded ALEPH data are quite correlated, as can be
seen from the upper panel of Fig.~\ref{ALEPH}, whose data is
much less scattered than the plotted errors would naively suggest.
The toy data samples exhibit points that are much
more scattered than the original data, as can be seen in
Fig.~\ref{ALEPH}, which shows the real data together with a
representative toy sample (both plotted with the same errors). The
strong correlations in the upper panel are not present in the toy
sample, which points to a problem in the posted correlation
matrix. According to the authors of the original analysis, the
publicly available data do not contain correlations due to
unfolding~\cite{OrsayGroup}.

The effect of the missing correlations seems to be small for the
doubly or triply pinched weights used in recent analyses based on the
publicly available ALEPH spectral data.\footnote{An exploratory
  reanalysis by the authors of Ref. [7] based on one exclusive channel
  ($\tau^-\to\pi^-\pi^0\nu_\tau$) suggests that this is indeed the
  case~\cite{OrsayGroup}.} More pronounced effects, however, may be
expected for FESRs based on unpinched weights, which are required for
a reliable exploration of DVs. A precise quantification awaits the
completion of a reanalysis of the ALEPH covariance matrices.

\section*{Acknowledgements}

{\ni We thank C.~Bernard, S. Descotes-Genon, M.~Mart\'inez, S.~Menke
  and R.~Miquel for discussions and M.~Davier, A.~H\"ocker,
  B.~Malaescu, and Z.~Zhang for correspondence. This work was
  supported in part by the Spanish Ministry (grants CICYT-FEDER
  FPA2007-60323, FPA2008-01430, CPAN CSD2007-00042), by the Catalan
  Government (grant SGR2009-00894), EU Contract MRTN-CT-2006-035482,
  NERSC (Canada), and the US Department of Energy.}


\begin{thebibliography}{99}

\bibitem{BNP1992}{E.~Braaten}, {S.~Narison}, and {A.~Pich},
%\newblock {QCD} analysis of the $\tau$ hadronic width,
 Nucl. Phys. B {\bf 373} (1992) 581.  
 %%CITATION = PHRVA,D58,096014;%%

\bibitem{ALEPH}S.~Schael {\it et al.}  [ALEPH Collaboration],
  %``Branching ratios and spectral functions of tau decays: Final ALEPH
  %measurements and physics implications,''
  Phys.\ Rept.\  {\bf 421} (2005) 191
  [arXiv:hep-ex/0506072].
  %%CITATION = PRPLC,421,191;%%

\bibitem{PichetalDVs}  See \eg~M.~Davier, L.~Girlanda, A.~H\"ocker and J.~Stern,
  %``Finite energy chiral sum rules and tau spectral functions,''
  Phys.\ Rev.\  D {\bf 58}, 096014 (1998)
  [arXiv:hep-ph/9802447];
M.~Gonzalez-Alonso, A.~Pich and J.~Prades,
  %``Violation of Quark-Hadron Duality and Spectral Chiral Moments in QCD,''
  Phys.\ Rev.\  D {\bf 81} (2010) 074007
  [arXiv:1001.2269 [hep-ph]]; Phys.\ Rev.\  D {\bf 82} (2010) 014019
  [arXiv:1004.4987 [hep-ph]].
  %%CITATION = PHRVA,D82,014019;%%

\bibitem{Vus1} See \eg~D.~Asner {\it et al.}, [HFAG]
  %``Averages of b-hadron, c-hadron, and tau-lepton Properties,''
    [arXiv:1010.1589 [hep-ex]];
    E.~Gamiz {\em et al.},
  %``V_us and m_s from hadronic tau decays,''
  Phys.\ Rev.\ Lett.\  {\bf 94} (2005) 011803
  [arXiv:hep-ph/0408044];  %%CITATION = PRLTA,94,011803;%%; 
K.~Maltman and C.~E.~Wolfe,
  %``Joint extraction of m(s) and V(us) from hadronic tau decays,''
  Phys.\ Lett.\  B {\bf 650} (2007) 27
  [arXiv:hep-ph/0701037].
  %%CITATION = PHLTA,B650,27;%%

\bibitem{OPAL}K.~Ackerstaff {\it et al.}  [OPAL Collaboration],
  %``Measurement of the strong coupling constant alpha(s) and the vector  and
  %axial-vector spectral functions in hadronic tau decays,''
  Eur.\ Phys.\ J.\  C {\bf 7} (1999) 571
  [arXiv:hep-ex/9808019].
  %%CITATION = EPHJA,C7,571;%%

\bibitem{Baikovetal2008} P.~A.~Baikov, K.~G.~Chetyrkin and J.~H.~K\"uhn,
  %``Order $\alpha^4_s$ QCD Corrections to $Z$ and $\tau$ Decays,''
  Phys.\ Rev.\ Lett.\  {\bf 101} (2008) 012002, 
  [arXiv:0801.1821 [hep-ph]].
  %%CITATION = PRLTA,101,012002;%%

\bibitem{Davieretal2008}{ M.~Davier} {\em et al.},
%\newblock {The determination of $\alpha_s$ from $\tau$ decays revisited},
\newblock { Eur. Phys. J.} {\bf C56}  (2008) 305, [arXiv:0803.0979 [hep-ph]].

\bibitem{BenekeJamin2008}  M.~Beneke and M.~Jamin,
  %``$\alpha_s$ and the $\tau$ hadronic width: fixed-order, contour-improved and
  %higher-order perturbation theory,''
  JHEP {\bf 0809}, 044  (2008),   [arXiv:0806.3156 [hep-ph]].
  %%CITATION = JHEPA,0809,044;%%

\bibitem{MaltmanYavin2008} K.~Maltman and T.~Yavin,
  %``Alpha_s(M_Z) from hadronic tau decays,''
  Phys.\ Rev.\  D {\bf 78} (2008) 094020,
  [arXiv:0807.0650 [hep-ph]].
  %%CITATION = PHRVA,D78,094020;%%

\bibitem{LatticeAlphaS} C.~McNeile {\em et al.},
  %``High-Precision c and b Masses, and QCD Coupling from Current-Current
  %Correlators in Lattice and Continuum QCD,''
  Phys.\ Rev.\  D {\bf 82} (2010) 034512
  [arXiv:1004.4285 [hep-lat]].
  %%CITATION = PHRVA,D82,034512;%%

\bibitem{Pivovarov}  A.~A.~Pivovarov,
  %``Renormalization group analysis of the tau-lepton decay within QCD,''
  Z.\ Phys.\  C {\bf 53} (1992) 461
%  [Sov.\ J.\ Nucl.\ Phys.\  {\bf 54} (1991) 676]
%  [Yad.\ Fiz.\  {\bf 54} (1991) 1114]
  [arXiv:hep-ph/0302003].
  %%CITATION = YAFIA,54,1114;%%
%
%\bibitem{PichLeDiberder}  
  F.~Le Diberder, A.~Pich,
  %``Testing QCD with tau decays,''
  Phys.\ Lett.\  {\bf B289 } (1992)  165-175.

\bibitem{CapriniFischer}
  I.~Caprini and J.~Fischer,
  %``$\alpha_s$ from $\tau$ decays: contour-improved versus fixed-order
  %summation in a new QCD perturbation expansion,''
  Eur.\ Phys.\ J.\  C {\bf 64}, 35 (2009)
  [arXiv:0906.5211 [hep-ph]].
  %%CITATION = EPHJA,C64,35;%%

\bibitem{Shifman} B.~Blok, M.~A.~Shifman and D.~X.~Zhang,
  %``An illustrative example of how quark-hadron duality might work,''
  Phys.\ Rev.\  D {\bf 57} (1998) 2691
  [Erratum-ibid.\  D {\bf 59} (1999) 019901]
  [arXiv:hep-ph/9709333].
  %%CITATION = PHRVA,D57,2691;%%

\bibitem{CGP2005}O.~Cat\`a, M.~Golterman, S.~Peris,
  %``Duality violations and spectral sum rules,''
  JHEP {\bf 0508}, 076 (2005), [hep-ph/0506004].

\bibitem{CGP2008} O.~Cat\`a, M.~Golterman, S.~Peris,
  %``Unraveling duality violations in hadronic tau decays,''
  Phys.\ Rev.\  {\bf D77 } (2008)  093006,  [arXiv:0803.0246 [hep-ph]]; Phys.\ Rev.\  {\bf D79 } (2009)  053002, [arXiv:0812.2285 [hep-ph]].

\bibitem{km98}
  K.~Maltman,
  %``Constraints on hadronic spectral functions from continuous families of
  %finite energy sum rules,''
  Phys.\ Lett.\  B {\bf 440}, 367 (1998)
  [arXiv:hep-ph/9901239].
  %%CITATION = PHLTA,B440,367;%%
%
%\bibitem{ds99}
  C.~A.~Dominguez and K.~Schilcher,
  %``Chiral sum rules and duality in {QCD},''
  Phys.\ Lett.\  B {\bf 448}, 93 (1999)
  [arXiv:hep-ph/9811261].
  %%CITATION = PHLTA,B448,93;%%

\bibitem{MILC}
  C.~Bernard {\it et al.}  [MILC Collaboration],
  %``Lattice calculation of heavy-light decay constants with two flavors of
  %dynamical quarks,''
  Phys.\ Rev.\  D {\bf 66}, 094501 (2002)
  [arXiv:hep-lat/0206016].
  %%CITATION = PHRVA,D66,094501;%%

\bibitem{Bohmetal} G.~Bohm, G.~Zech, {\it Introduction to Statistics and Data
Analysis for Physicists}, Verlag DESY, ISBN 978-3-935702-41-6.

\bibitem{4pi} 
  M.~Davier, S.~Eidelman, A.~H\"ocker and Z.~Zhang,
  %``Confronting spectral functions from e+ e- annihilation and tau decays:
  %Consequences for the muon magnetic moment,''
  Eur.\ Phys.\ J.\  C {\bf 27}, 497 (2003)
  [arXiv:hep-ph/0208177];
  %%CITATION = EPHJA,C27,497;%%
  V.~P.~Druzhinin,
  %``Study of e+e- annihilation at low energies,''
  arXiv:0710.3455 [hep-ex].
  %%CITATION = ARXIV:0710.3455;%%

\bibitem{ALEPHTables} {\tt aleph.web.lal.in2p3.fr/tau/specfun.html}
 
\bibitem{OrsayGroup} M.~Davier, A.~H\"ocker, B.~Malaescu and Z.~Zhang, private communication.

\end{thebibliography}
\end{document}